\documentclass{ws-procs9x6}

\begin{document}


\markboth{J. P. Draayer, \emph{et al}.}{Exactly Solvable Pairing Models}


%

\catchline{}{}{}{}{}

%


\title{EXACTLY SOLVABLE PAIRING MODELS}
\author{\footnotesize \underline{J. P. DRAAYER}, V. G. GUEORGUIEV, K. D.
SVIRATCHEVA, C. BAHRI}
\address{Department of Physics and Astronomy, Louisiana State University, \\
Baton Rouge, LA 70803, USA}
\author{ \footnotesize FENG PAN}
\address{Department of Physics, Liaoning Normal University, \\
Dalian, 116029, P. R. China}
\author{ \footnotesize A. I. GEORGIEVA}
\address{Institute of Nuclear Research and Nuclear Energy,\\
Bulgarian Academy of Sciences, Sofia, Bulgaria}
\maketitle

\centerline{Abstract}
Some results for two distinct but complementary exactly solvable algebraic
models for pairing in atomic nuclei are presented: 1) binding energy 
predictions
for isotopic chains of nuclei based on an extended pairing model that includes
multi-pair excitations; and 2) fine structure effects among excited 
$0^+$ states
in $N \approx Z$ nuclei that track with the proton-neutron ($pn$) and
like-particle isovector pairing interactions as realized within an algebraic
$sp(4)$ shell model. The results show that these models can be used 
to reproduce
significant ranges of known experimental data, and in so doing, confirm their
power to predict pairing-dominated phenomena in domains where data is
unavailable.

\section{Introduction}

Pairing is an important interaction that is widely used in nuclear and other
branches of physics. In this contribution we present some results that follow
from exact algebraic solutions of an extended pairing model that includes
multi-pair excitations and that is designed to reproduce binding energies of
deformed nuclei,\cite{Feng'04} and the $sp(4)$ pairing model that can 
be used to
track fine structure effects in excited $0^+$ states in medium mass
nuclei.\cite{SGD03} The results show that these models can be used to reproduce
significant ranges of known experimental data, and in so doing, confirm their
power to predict pairing-like phenomena in domains where data is unavailable or
simply not well understood, such as binding energies for proton or neutron rich
nuclei far off the line of stability and the fine structure of proton-neutron
systems that are critical to understanding the $rp$-process in nucleosynthesis.

The Bardeen-Cooper-Schrieffer (BCS)\cite{BCS} and Hartree-Fock-Bogolyubov
(HFB)\cite{Ring-Schuck} methods for finding approximate solutions when pairing
plays an important role are well known. However, the limitations of 
BCS methods,
when applied in nuclear physics, are also well understood. First of all,  the
number of valence particles ($n\sim 10$) that dominate the behavior 
of low-lying
states is too few to support the underlying assumptions of the approximations,
that is, particle number fluctuations are non-negligible. As a result, particle
number-nonconservation effects can lead to serious problems such as spurious
states, nonorthogonal solutions, and so on. In addition, an essential feature
of pairing correlations are differences between neighboring even and odd mass
nuclei, which are driven mainly by Pauli blocking effects. It is difficult to
treat these even-odd differences with either the BCS or HFB theories because
different quasi-particle bases must be introduced for different blocked levels.
Another difficulty with approximate treatments of the pairing interaction is
related to the fact that both the BCS and the HFB approximations break down for
an important class of physical situations. A remedy that uses particle number
projection techniques complicates these methods and does not help achieve a
better description of higher-lying states.

\section{Mean-field plus Extended Pairing Model}

The importance of having exact solutions of the pairing Hamiltonian has driven
a great deal of work in recent years. In particular, building on Richardson's
early work\cite{Richarson} and extensions to it based on the Bethe ansatz,
several authors have introduced novel approaches.\cite{BetheAnsatz,Feng'99} For
the algebraic approaches based on the Bethe ansatz, the solutions are provided
by
a set of highly non-linear Bethe Ansatz Equations (BAE). Although these
applications demonstrate that the pairing problem is exactly 
solvable, solutions
are not easily obtained and normally require extensive numerical work,
especially
when the number of levels and valence pairs are large.  This limits the
applicability of the methodology to relatively small systems; in particular, it
cannot be applied to large systems such as well-deformed nuclei.

\subsection{Algebraic Underpinnings of the Theory}

The standard pairing Hamiltonian for well-deformed nuclei is given by
\begin{equation}
\hat{H}=\sum_{j=1}^{p}\epsilon _{j}n_{j}-G\sum_{i,j=1}^{p}a_{i}^{+}a_{j},
\label{eqno1}
\end{equation}
where $p$ is the total number of single-particle levels, $G>0$ is the pairing
strength, $\epsilon _{j}$ is single-particle energies taken for example from a
Nilsson model, $n_{j}=c_{j\uparrow }^{\dagger }c_{j\uparrow }+c_{j\downarrow
}^{\dagger }c_{j\downarrow }$ is the fermion number operator for the $j$-th
single particle level, and $a_{i}^{+}=c_{i\uparrow }^{\dagger }c_{i\downarrow
}^{\dagger }$ ($a_{i}=(a_{i}^{+})^{\dagger }=c_{i\downarrow }c_{i\uparrow }$)
are pair creation (annihilation) operators. The up and down arrows in these
expressions denote time-reversed states. Since each level can only be occupied
by one pair due to the Pauli Exclusion Principle, the Hamiltonian (\ref{eqno1})
is also
equivalent to a finite site hard-core Bose-Hubbard model with infinite range
one-pair hopping and infinite on-site repulsion. Specifically, the operators
$a_{i}^{+}$, $a_{i}$, and $n_{i}^{a}=n_{i}/2$ satisfy the following
hard-core boson algebra:
\begin{equation}
(a_{i}^{+})^{2}=0,~~[a_{i},a_{j}^{+}]=\delta
_{ij}(1-2n_{i}^{a}),~~[a_{i}^{+},a_{j}^{+}]=[a_{i},a_{j}]=0.
\label{eqno2}
\end{equation}

The extended pairing Hamiltonian adds multiple-pair excitations to the standard
pairing interaction (\ref{eqno1}):
\begin{equation}
\hat{H}=\sum_{j=1}^{p}\epsilon _{j}n_{j}-G\sum_{i,j=1}^{p}a_{i}^{+}a_{j}
-G \sum_{\mu =2}^{\infty }{\frac{1}{{{(\mu !)}^{2}}}}
\sum_{i_{1}\neq \cdots \neq i_{2\mu}}
a_{i_{1}}^{+}\cdots a_{i_{\mu }}^{+}a_{i_{\mu +1}}\cdots a_{i_{2\mu }},
\label{eqno3}
\end{equation}
where no pair of indices among the
$\{i_{1},i_{2},\cdots, i_{2\mu}\}$ are the same for
any $\mu$.
With this extension, the model is exactly solvable.\cite{Feng'04}  In
particular,
the $k$-pair excitation
energies of (\ref{eqno3}) are given by the expression:

\begin{equation}
E_{k}^{(\zeta )}=\frac{2}{x{^{(\zeta )}}}-G(k-1),
\label{eqno9}
\end{equation}
where the undetermined variable $x^{(\zeta )}$ satisfies
\begin{equation}
{\frac{2}{{x^{(\zeta )}}}}+\sum_{1\leq i_{1}<i_{2}<\cdots <i_{k}\leq p}
{\frac{G}{(1-{x^{(\zeta )}\sum_{\mu =1}^{k}\epsilon _{i_{\mu }})}}=0}.
\label{eqno10}
\end{equation}
The additional quantum number $\zeta $ can be understood as the
$\zeta$-th solution of (\ref{eqno10}). Similar results can be shown to hold for
even-odd systems except that the index $j$ of the level occupied by the single
nucleon should be excluded from the summation and the
single-particle energy term $\epsilon _{j}$ contributing to the 
eigenenergy from
the first term of (\ref{eqno3}) should be included. Extensions to many
broken-pair cases are straightforward. If (\ref{eqno10}) is rewritten in terms
of
a new variable $z^{(\zeta )}=2/[G x^{(\zeta )}]$ and the 
dimensionless energy of
a `grand' boson $\tilde{E}_{i_{1}i_{2}...i_{k}}=\sum_{\mu =1}^{k}\frac{2 {
\epsilon
}_{i_{\mu }}}{G},$ (\ref{eqno10}) reduces to:
\begin{equation}
1=\sum_{1\leq i_{1}<i_{2}<\cdots <i_{k}\leq p}{\frac{1}{(
\tilde{E}_{i_{1}i_{2}...i_{k}}{-}z^{(\zeta )}{)}}}.
\label{eqno11}
\end{equation}
Since there is only a single variable $z^{(\zeta)}$ in (\ref{eqno11}), the
zero points of the function can be determined graphically in a manner that
is similar to the one-pair solution of the TDA and RPA approximations with
separable potentials.\cite{Ring-Schuck}

\subsection{Application to the $^{154-181}$Yb Isotopes}

A study of the binding energies of well-deformed nuclei within the framework of
the extended pairing model is currently in progress.\cite{SecondPaper}
Typically,
the single-particle energies of each nucleus are calculated within the deformed
Nilsson shell model with deformation parameters taken from Moller and
Nix;\cite{Moller & Nix} experimental binding energies are taken from Audi, et
al;\cite{Audi G} and, theoretical binding energies are calculated
relative to a particular core. For an even number of neutrons, only pairs of
particles (bosons-like structures) are considered. For an odd number of
neutrons,
Pauli blocking of the Fermi level of the last unpaired fermion is envoked with
the remaining fermions are considered to be an even $A$ fermion system. Using
(\ref{eqno9}) and (\ref{eqno10}), values of $G$ are calculated so that the
experimental and theoretical binding energy match exactly. Note that 
for a given
set of single-particle energies there is an upper limit to the binding energy
for
which a physically meaningful exact solution can be constructed. This upper
value
on the binding energy is given by the energy of the lowest `grand' boson, with
energy given by $\sum_{\mu =1}^{k}2 {\epsilon }_{\mu}$.

\begin{figure}[htbp]
\centerline{\epsfxsize=4.5in\epsfbox{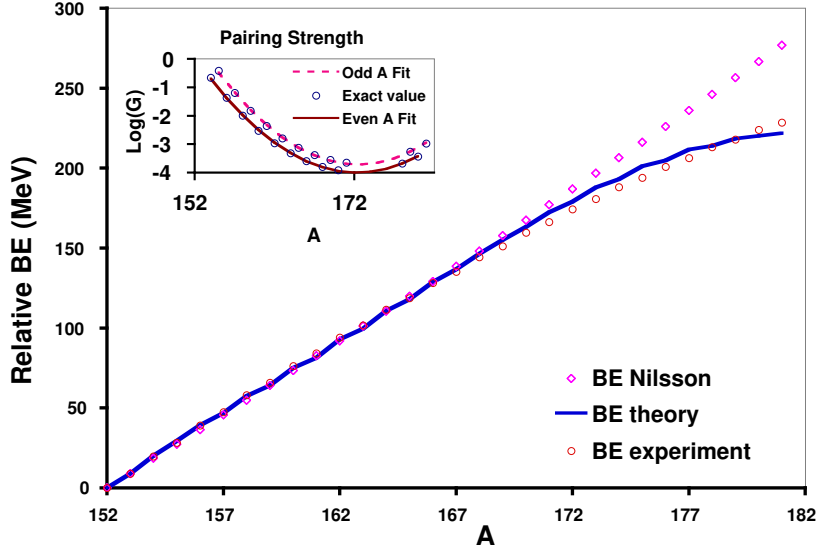}}
\vspace*{8pt}
\caption{The solid line gives the theoretical binding energies of the Yb
isotopes relative to that of the $^{152}$Yb core. The single-particle
energy scale is set from the binding energy of $^{153}$Yb. The inset
shows the fit to values of $G$ that reproduce exactly the experimental
data. The two fitting functions are:
$\log(G(A))=662.2247 - 7.7912 A + 0.0226 A^2$ for even values of $A$
and
$\log(G(A))=716.3279 - 8.4049 A + 0.0244 A^2$ for odd values of $A$.
The Nilsson BE energy is the lowest configuration energy of the non-interacting
system.}
\label{Yb-isotopes}
\end{figure}

As a first application of the theory,  we calculated the binding energies for
the $^{154-181}$Yb isotopes and extracted the corresponding $\log(G)$ 
values for
the extended pairing model. The binding energy of the closed neutron shell
nucleus $^{152}$Yb was taken to be the zero-energy reference point. 
Its odd-$A$
$^{153}$Yb neighbor was assumed to be well described by the independent
particle model with Nilsson single-particle energies; this means that the
pairing interaction terms have no affect on $^{153}$Yb. The energy scale
applied to the Nilsson single-particle energies, which is 3/4 for pure harmonic
oscillator interaction, was set so that the binding energy of $^{153}$Yb is
reproduced by the independent particle model.\cite{Ring-Schuck} For all the
other
nuclei we solved for the pairing strength $G(A)$ that reproduces the
experimental
binding energies exactly within the selected model space, the latter consisting
of the neutron single-particle levels between the closed shells with magic
numbers
50 and 82. The structure of the model space is reflected in the values of
$G(A)$.
In particular, $\log(G(A))$ has a smooth quadratic behavior for even- and
odd-$A$
values with a minimum in the middle of the model space where the size of the
space
is a maximal. As shown in Figure \ref{Yb-isotopes}, although the even- and
odd-$A$
curves are very similar, they are shifted from one another due to the even-odd
mass difference.

To summarize, in this section we reviewed the extended pairing model and tested
its predictive power using the $^{172-177}$Yb isotopic chain as an example. In
particular, calculations of the pairing strength $G$ were carried out for the
$^{154-171}$Yb and $^{178-181}$Yb isotopes but not for the $^{172-177}$Yb
isotopes
that are in the  middle of the model space where the computations are more
involved. The even- and odd-$A$ $\log(G)$ curves, which were assumed to have a
quadratic polynomial form and therefore determined by three 
parameters, were fit
to the two date sets which consist of 11 data points each, one for the even-$A$
isotopes and another for odd $A$.  From the quadratic polynomial fit to the
$\log(G)$ values, we then calculate the theoretical values of the binding
energy
for all the nuclei shown in Figure \ref{Yb-isotopes}. The prediction is very
good when compared to the experimental numbers. Thus, based on 
experimental data
of the nuclei in the upper and lower parts of the shell and an 
assumed quadratic
from for $\log(G)$ that was fit to this data, we were able to make reasonable
estimates for the binding energies of mid-shell nuclei. Based on this simple
exercise, we conclude that the extended pairing model has good predictive power
for binding energies. Indeed, this early success suggests that the extended
pairing model may have broader applicability to other well-deformed nuclei as
well as other physical systems where pairing plays an important role.

\section{Algebraic $sp(4)$ Pairing Model}

The recent renaissance of studies on pairing is related to the search of a
reliable microscopic theory for a description of medium nuclei around
the $N=Z$ line, where like-particle pairing comprises only a part of the
complicated nuclear interaction in this region. This is because for such nuclei
protons and neutrons occupy the same major shells and their mutual interactions
are expected to influence significantly the structure and decay of 
these nuclei.
Such a microscopic framework is as well essential for astrophysical
applications,
for example the description of the $rp$-process in nucleosynthesis, which runs
close to the proton-rich side of the valley of stability through reaction
sequences of proton captures and competing $\beta $ decays.\cite{Langanke98S}
The revival of interest in pairing correlations is also prompted by the
initiation of radioactive beam experiments, which advance towards 
exploration of
`exotic' nuclei, such as neutron-deficient or $N\approx Z$ nuclei far off the
valley of stability.

In our search for a  microscopic description of pairing in the broad range of
nuclei with mass numbers $32\le A\le 100$ with protons and neutrons filling the
same major shell, we employ an $sp(4)$ algebraic model that accounts for
proton-neutron and like-particle pairing correlations and higher-$J$
proton-neutron interactions, including the so-called symmetry and Wigner
energies.\cite{SGD03} The nuclei classified within a major shell 
possess a clear
$Sp(4)$ dynamical symmetry. The basis operators of the $sp(4)$ algebra ($\sim
so(5)$\cite{HelmersHechtG,EngelLV96}) have a distinct physical meaning:
$N _{\pm 1}$ counts the total number of protons (neutrons) (and hence
$\hat{N}=N_{+1}+N_{-1}$ is the total number operator), the operators
$T_{0, \pm }$  are
related to isospin (where  $T_0=(N_{+1}-N_{-1})/2$ is the third projection of
isospin), while the six operators $A^{\dagger} _{-1,0,1}$ ($A_{-1,0,1}$) create
(annihilate) a pair of total angular momentum $J^{\pi}=0^+$ and isospin $T=1$.
The
model Hamiltonian with an $Sp(4)$ dynamical symmetry,
\begin{eqnarray}
H =&-G\sum _{i=-1}^{1}A^{\dagger }_{i}
A_{i}-F A^{\dagger }_{0}A_{0}-\frac{E}{2\Omega} (T
^2-\frac{3\hat{N}}{4 })
\nonumber \\
&-D(T 
_{0}^2-\frac{\hat{N}}{4})-C\frac{\hat{N}(\hat{N}-1)}{2}-\epsilon 
\hat{N},
\label{clH}
\end{eqnarray}
includes a two-body isovector ($T=1$) pairing interaction and a
diagonal isoscalar ($T=0$) force, which is proportional to a symmetry 
and Wigner
term ($T(T+1)$-like dependence). In addition, the  $D$-term introduces isospin
symmetry breaking and the $F$-term accounts for a plausible, still extremely
weak, isospin mixing. This Hamiltonian conserves the number of particles ($N$),
the third projection of isospin ($T_0$) and angular momentum, and changes the
like-particle seniority quantum number by zero or $\pm 2$, the latter implies
scattering of a $pp$ pair and  a $nn$ pair into two $pn$ pairs and vice versa.
The interaction strength parameters in (\ref{clH}) are estimated in 
optimum fits
to the  lowest isobaric analog $0^+$ state experimental energies of total of
$149$ nuclei\cite{SGD03} and are found to have a smooth dependence on the
nuclear
mass $A$,
\begin{eqnarray}
&\frac{G}{\Omega }=\frac{23.9 \pm 1.1}{A}, \quad
\frac{E}{2\Omega }=\frac{-52 \pm 5}{A},\nonumber \\
&D=\frac{-37 \pm 5}{A}+(-0.24\pm 0.09), \quad C=\left( \frac{32 \pm 1}{A}
\right)^{1.7\pm 0.2},
\end{eqnarray}
where $2\Omega =\Sigma _j (2j+1)$ is the shell dimension.

The basis states are constructed as ($T=1$)-paired fermions,
$\left| n_{1},n_{0},n_{-1}\right) =\left( A^{\dagger  }_{1} \right)
^{n_{1}}\left( A^{\dagger  }_{0}\right) ^{n_{0}}\left( A^{\dagger 
}_{-1}\right)
^{n_{-1}}\left| 0\right\rangle ,
$
and model the $0^+$ ground state for even-even and some odd-odd nuclei and the
corresponding isobaric analog  excited $0^+$ state for even-$A$ nuclei in
a significant range of nuclei, $32\leq A \leq 100$.  The properties of these
states are described well by the $Sp(4)$ dynamical symmetry model, including
quite good agreement of the isobaric analog $0^+$ state energy spectra with
experiment, and in addition the remarkable reproduction of their detailed
structure properties.

\subsection {Energy Spectra of Isobaric Analog $0^+$ States}

\begin{figure}[th]
\centerline{\epsfxsize=4.5in\epsfbox{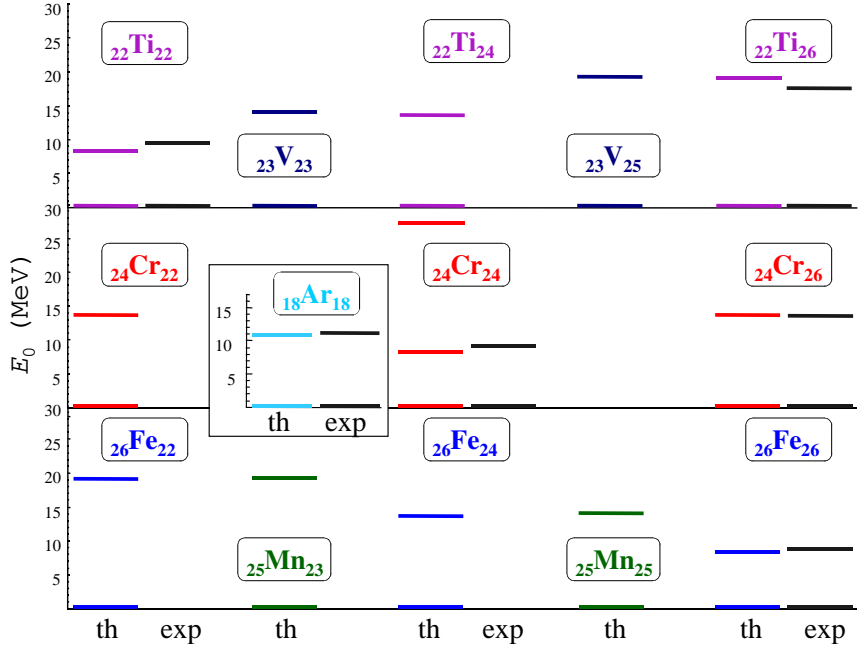}}
\caption{Theoretical (`th') and experimental (`exp') energy spectra of the
higher-lying isobaric analog $0^+$ states for isotopes in $1f_{7/2}$ (in
$1d_{3/2}$ (insert)).}
\label{enSpectraCa}
\end{figure}

The $Sp(4)$ model leads to a very good reproduction of the experimental
energies of the lowest isobaric analog $0^+$ state for even-$A$ 
nuclei (that is,
binding energies for even-even and some odd-odd nuclei) with nuclear masses
$32\leq A \leq 100$.\cite{SGD03} This result follows from the very small
deviation  (estimated by the $\chi$-statistics) between experimental
energies and the corresponding theoretical energies predicted in optimization
procedures, namely $\chi =0.496 $ in the $1d_{3/2}$ shell, $\chi 
=0.732 $ in the
$1f_{7/2}$ shell and $\chi =1.787 $ in the
$1f_{5/2}2p_{1/2}2p_{3/2}1g_{9/2}$ major shell. Without varying the values of
the
interaction strength parameters, the energy of the higher-lying isobaric analog
$0^+$ states can be theoretically calculated and they agree 
remarkably well with
the
available  experimental  values for the single-$j$
$1d_{3/2}$ and $1f_{7/2}$ orbits (Figure \ref{enSpectraCa}). However, such a
comparison to experiment is impossible for the nuclei in the region
with nuclear masses $56<A<100$, since their energy spectra are not yet
completely
measured, especially the higher-lying $0^+$ states.

The agreement, which is observed throughout both single-$j$ shells, represents
an important result. This is because the higher-lying isobaric analog
$0^+$ states constitute an experimental set independent of the data that
determines the interaction strength parameters in (\ref{clH}). 
Therefore, such a result
is, first, an independent test of the physical validity of the 
strength parameters,
and, second, an indication that the interactions interpreted by the model
Hamiltonian are the main driving force that defines the properties of these
states.  In this way, the simple $Sp(4)$ model provides for a reasonable
prediction of the isobaric analog (ground and/or excited) $0^+$ states in
proton-rich nuclei with energy spectra not yet  experimentally fully explored.
For example, in the case of the $1f_{7/2}$ level the binding energy of the
proton-rich $^{48}$Ni nucleus is estimated to be $E_0=348.19$ MeV, which is
$0.07\%$ greater than the sophisticated semi-empirical estimate of Moller and
Nix.\cite{Moller & Nix} Likewise, for the odd-odd nuclei that do not have
measured energy spectra the theory  can predict the energy of their lowest
$0^{+}$ isobaric analog state: $358.62$ MeV ($^{44}$V), $359.34$ MeV
($^{46}$Mn),
$357.49$ MeV ($^{48}$Co), $394.20$ MeV ($^{50}$Co). The $Sp(4)$ model predicts
the
relevant $0^+$ state energies for additional 165 even-$A$ nuclei in the medium
mass region of the
$1f_{5/2}2p_{1/2}2p_{3/2}1g_{9/2}$ major shell. The binding energies for 25 of
them are also calculated in Moller and Nix.\cite{Moller & Nix} For these
even-even nuclei, we predict binding energies that on average are by $0.05\%$
less than the semi-empirical approximation.\cite{Moller & Nix}

\subsection {N = Z Irregularities, Staggering and the Pairing Gap}

\begin{figure}[th]
\centerline{\epsfxsize=4.5in\epsfbox{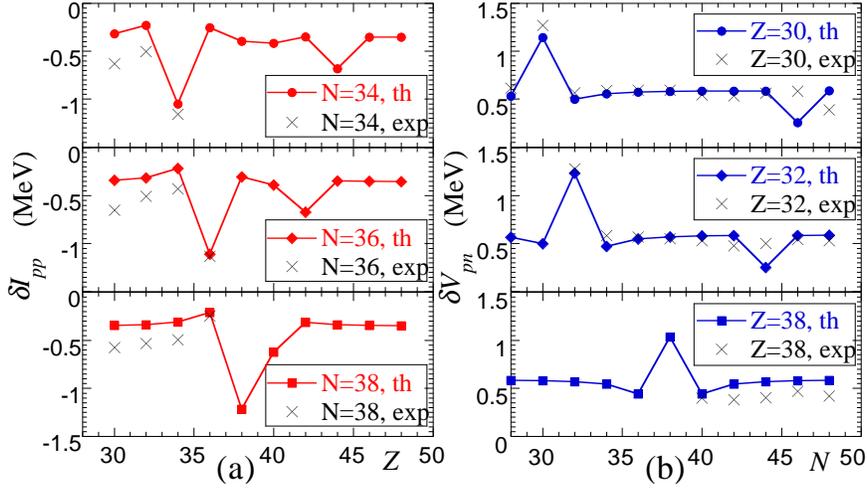}}
\caption{Second discrete derivatives of the energy function $E_0$: (a) $\delta
I_{pp(nn)}(N_{\pm 1 })=\frac{E_0(N_{\pm 1}+2)-2E_0(N_{\pm 1})+E_0(N_{\pm
1}-2)}{4}$ versus $N_{\pm 1}$, as an estimate for the  non-pairing 
like-particle
nuclear interaction in MeV for the  $N(Z)=34,36,38$-multiplets; (b) $\delta
V_{pn}(N_{+1},N_{-1})=
\frac{E_0(N_{+1}+2,N_{-1}+2)-E_0(N_{+1}+2,N_{-1})-
E_0(N_{+1},N_{-1}+2)+E_0(N_{+1},N_{-1})}{4}$ versus $N_{+1}$ and 
$N_{-1}$, as an
estimate for the residual interaction between the last proton and the last
neutron in MeV for Zn, Ge, Sr isotopes.}
\label{NiDer2NpVpn}
\end{figure}

The theoretical $Sp(4)$ model can be further tested through second- and
higher-order discrete derivatives of the energies of the lowest isobaric analog
$0^+$ states in the $Sp(4)$ systematics, without any parameter variation. The
theoretical discrete derivatives under investigation not only follow the
experimental patterns but their magnitude was found to be in a remarkable
agreement with the data. The proposed model has been used to successfully
interpret: the two-proton (two-neutron)  separation energy $S_{2p(2n)}$ for
even-even nuclei (hence determined the two-proton drip line), the  $S_{pn}$
energy difference when a $pn$ $T =1$ pair is added, the
observed\cite{Brenner90ZamfirC91} irregularities around $N=Z$ (Figure
\ref{NiDer2NpVpn}), the like-particle and $pn$ isovector pairing gaps, and the
prominent ``$ee$-$oo$" staggering between even-even and odd-odd nuclides. We
suggest that the oscillating ``$ee$-$oo$" effects correlate with the 
alternating
of the seniority  numbers related to the $pn$ and like-particle  isovector
pairing, which is in addition to the larger contribution due to the
discontinuous change in isospin values associated with the symmetry
energy.\cite{SGD03stg}

The present study brings forward a very useful result. We find a finite energy
difference of the energy function $E_0$,
\begin{eqnarray} &&E_0(N,T_0+1)-2E_0(N,T_0)+E_0(N,T_0-1)= \nonumber
  \\
&&=E_0(N_{+1}+1,N_{-1}-1)-2E_0(N_{+1},N_{-1}) + E_0(N_{+1}-1,N_{-1}+1),
\label{Stg2i}
\end{eqnarray}
that, for the specific case $T_0=0$ (or $N=Z$), can be interpreted as an
isovector pairing gap, $\tilde{\Delta }=\Delta_{pp}+\Delta_{nn}-2\Delta_{pn}$,
which is related to the like-particle and $pn$ isovector pairing gaps. Indeed,
they correspond to the $T=1$ pairing mode because we do not consider 
the binding
energies for all the nuclei  but the respective isobaric analog $0^+$ 
states for
the odd-odd nuclei with a $J\neq 0^+$ ground state. This investigation is the
first of its kind. Moreover, the relevant energies are corrected  for the
Coulomb
interaction and therefore the isolated effects reflect solely the nature of the
nuclear interaction. In addition, the discrete derivative filter (\ref{Stg2i})
can be used to estimate the pairing gaps for all the nuclei within a 
major shell
when only the contribution of the pairing energy is considered in the $E_0$
energy
function. In this way, the like-particle pairing gap is found to be in a very
good agreement with the $12/\sqrt{A}$ experimental
approximation.\cite{BohrMottelson} Small deviations from the experimental data
are  attributed to other two-body interactions or higher-order 
correlations that
are not included in the theoretical model.

In summary, the symplectic $Sp(4)$ scheme allows not only for an extensive
systematic study of various experimental patterns of the even-$A$ nuclei, it
also offers a simple $sp(4)$ algebraic model for interpreting the results and
predicting properties of nuclei that are not yet experimentally explored.  The
outcome of the present investigation shows that, in comparison to  experiment,
the $sp(4)$ algebraic approach reproduces not only overall trends of the
relevant
energies but as  well the smaller fine features driven by isovector pairing
correlations and higher-$J$ $pn$  and like-particle nuclear interactions.

\section{Conclusion}

Results for two distinct but complementary exactly solvable algebraic models
for pairing in atomic nuclei have been presented: 1) binding energy predictions
for isotopic chains of nuclei based on an extended pairing model that includes
multi-pair excitations; and 2) fine structure effects among excited 
$0^+$ states
in $N \approx Z$ nuclei that track with the proton-neutron ($pn$) and
like-particle isovector pairing interactions as realized within an algebraic
$sp(4)$ shell model. The results show that both models can be used to reproduce
significant ranges of known experimental data, and in so doing confirm their
power to predict pairing-dominated phenomena in domains where data is either
not,
or only partially available or simply not well understood in terms of 
applicable
models.

In addition, it is important to reiterate that both approaches, the extended
pairing model and the algebraic $sp(4)$ model, yield exact analytic 
solutions to
their respective pairing problems. As the examples show, this is important for
applications, but it is also important for theory as having exact solutions
available gives one an opportunity to test approximate and perhaps simpler to
apply approaches, such as the BCS scheme. Other limits as well as extensions of
these theories are under investigation.

\section*{Acknowledgments}
Support from the U.S. National Science Foundation (0140300), the 
Natural Science
Foundation of China (10175031), and the Education Department of Liaoning
Province
(202122024) is acknowledged.

\end{document}